\documentclass[twocolumn,superscriptaddress,aps,preprintnumbers,amsmath,amssymb,sort&compress,nofootinbib
,a4paper
,prl
]{revtex4-2}
  
\usepackage{amsfonts,amsmath,amssymb,ascmac,bm,tensor}
\usepackage{fnpct} 
\usepackage{comment}
\usepackage{ifpdf}
\usepackage{slashed}
\usepackage{color}
\usepackage[mathscr]{eucal}
\usepackage{fourier}
\usepackage[utf8]{inputenc}
\usepackage{physics}
\usepackage{cancel}
\usepackage{subfig}
\ifpdf
  \usepackage{graphicx}     %   usepackage without driver option
  \usepackage[bookmarksopen,colorlinks=true,linkcolor=bblue,citecolor=bblue,urlcolor=ppink]{hyperref}
\else     % For (p)LaTeX + dvipdfmx
\fi

\usepackage{cancel}
\usepackage{ulem}
\usepackage{amsfonts}
\usepackage{amssymb}
\usepackage{graphicx}
\usepackage{amsmath,bm}
\usepackage{enumerate}
\usepackage{mathtools}
\usepackage{tikz} \usetikzlibrary{calc}
\usepackage{subfloat}
\usepackage{subfig}
\usepackage{xcolor}
\usepackage{float}
\usepackage{color}
\usepackage{here}

\usepackage{mathrsfs}
\usepackage{float,epsfig}
\usepackage{dcolumn}% Align table columns on decimal point
\usepackage{pgfplots}
\usepackage{graphicx}% Include figure files
\usepackage{bm}% bold math
\usepackage{amsmath,amssymb,amsthm}
\definecolor{red}{rgb}{1,0,0}
\definecolor{darkred}{rgb}{0.6,0,0}
\definecolor{darkgreen}{rgb}{0.992447,0.623778,0.034597}
\definecolor{ppink}{rgb}{1,0.4,0.4}
\definecolor{bblue}{rgb}{0.284602,0.317763,0.963947}
\definecolor{purple}{rgb}{0.5 ,0, 0.7}

\makeatletter
\newcommand\footnoteref[1]{\protected@xdef\@thefnmark{\ref{#1}}\@footnotemark}
\makeatother
 
\allowdisplaybreaks[1]

\definecolor{lime}{HTML}{A6CE39}
\newcommand{\orcidicon}{%
    \begin{tikzpicture}
    \draw[lime, fill=lime] (0,0)
        circle [radius=0.16]
        node[white] {{\fontfamily{qag}\selectfont \tiny ID}};
    \draw[white, fill=white] (-0.0625,0.095)
        circle [radius=0.007];
    \end{tikzpicture}   \hspace{-2mm}
}
\newcommand\orcidKarima{{\href{https://orcid.org/0000-0001-5419-8516}{\orcidicon}}}
\newcommand\orcidJamal{{\href{https://orcid.org/0000-0002-4463-4203}{\orcidicon}}}
\newcommand\orcidHasan{{\href{https://orcid.org/0000-0001-7408-0910}{\orcidicon}}}

\begin{document}

%%%%%%%%%%%%%%%%%%%%%%%%%%%
%%%%%%%%%%% Title %%%%%%%%%%%
%%%%%%%%%%%%%%%%%%%%%%%%%%%

\title{
  Evidence for NO violation of the second law in extended black hole thermodynamics}
\author{H. El Moumni \orcidHasan}	
\email{h.elmoumni@uiz.ac.ma (Corresponding author)}
\author{J. Khalloufi \orcidJamal}
\email{jamalkhalloufi@gmail.com}
\affiliation{\small LPTHE, Physics Department, Faculty of Sciences,  Ibn Zohr University, Agadir, Morocco.}

\author{K.  Masmar \orcidKarima.}
\email{karima.masmar@gmail.com}
\affiliation{\small Laboratory of  High Energy Physics and Condensed Matter
HASSAN II University, Faculty of Sciences Ain Chock, Casablanca, Morocco. }

\begin{abstract}
%\cite{Dai:2021dog}The purpose of the current letter is to give some relevant clarification into the
 The purpose of the current letter is to give some relevant clarification on the validity of the laws of thermodynamics and the stability of the horizon by scalar ﬁeld scattering formalism. On the one side, the connection between the energy of the absorbed particle and the change of the enthalpy of the black hole appears to resolve the violation of the second law. On the other side, such connection stipulates a fixed rank of the gauge group $N$ in the boundary conformal field theory which is against the extended phase space spirit, where the cosmological constant is allowed to vary inducing a holographical dually changing of $N$. By recalling the Grand potential, we suggest more stringent conditions under which the second law holds taking into account the missing information about the variation of the cosmological constant ie. the pressure.
Our result offers direct evidence of no violation of the second law in the extended phase space.
\end{abstract}

\date{\today}
\maketitle
%\preprint{CERN-TH-2020-182}
%\preprint{DESY 20-188}

%\emph{Introduction.}---

The second law of thermodynamics is one of  the most crucial  principles in the foundation of physics, so much so that Sir Arthur
Eddington once wrote the according \cite{Eddington1935-EDDNPI}: \\
{\it  “The law that
entropy always increases holds, I think, the supreme
position among the laws of Nature. If someone points
out to you that your pet theory of the universe is in
disagreement with Maxwell’s equations—then so much
the worse for Maxwell’s equations. If it is found to be
contradicted by observation—well, these experimentalists
do bungle things sometimes. But if your theory is found to
be against the Second Law of Thermodynamics I can give
you no hope; there is nothing for it but to collapse in
deepest humiliation.”}\\
With a deep conviction in such assertion, we undertake what we think is a piece of irrefutable evidence that the second law of thermodynamics can't be violated and it's the ultimate principle governing our universe.   Rigorously,   numerous papers have claimed that the horizon area -and thus the entropy- of near extremal black holes in AdS spacetimes can be reduced by dropping particles into them \cite{art7,art8, art9}. Such papers are based on  Ref.\cite{Gwak:2019rcz} assumption where Gwak relates the energy flux to the internal energy: $dE=dU$. Lately, in Ref.\cite{Hu:2019lcy}, the authors bring up that the violation of the second law is a direct consequence of an underlying incorrect assumption and they suggest that the energy flux of the scalar ﬁeld is assumed to change the black hole’s enthalpy $dE=dM$ rather than the internal energy.
  A fundamental step associated with this analysis is based on what we think is an insufficient argument:  Assuming that in the standard holography, one works with the large $N$ regime in the CFT, but this $N$, however large, {\bf is fixed} is against the essence of the extended phase space formalism where the cosmological constant is allowed to vary \cite{Henneaux:1984ji,Kubiznak:2012wp,art20,Wei:2015iwa}. % {\bf giving rise to  $N$ variation}.  
  Additionally, associating the holographic dictionary to the extended phase space framework has led to a rich black hole phase structure in the string theory framework \cite{Zhang:2014uoa,Belhaj:2015uwa,Chabab:2015ytz}, has unveiled the new notion of the holographic heat engines \cite{Johnson:2014yja} and opened new windows in the AdS/CFT correspondence investigations.
 %\textcolor{red}{$\bigstar$}
 
In the extended phase space, where the cosmological constant $\Lambda$ is viewed as a dynamic variable, the pressure $P=-\frac{\Lambda}{8\pi}=\frac{3}{8\pi\ell^2}$ is introduced as a thermodynamic quantity of black hole and its conjugate quantity $V$ denoting the thermodynamical volume. 
The first law of thermodynamics determines the infinitesimal change of the  mass  as \cite{art2, art3}
\begin{equation}\label{9}
dM = T dS + V dP + \Phi dQ,
\end{equation}
where $T$ is the black hole temperature, $S$ denotes the entropy, $\Phi$ stands for the electric potential, and $Q$ represents the electric charge.

%One consider the motion of charged particle in RN-AdS black hole background,
%\begin{equation}
%f(r)=1-\frac{2M}{r}+\frac{Q^2}{r^2}+\frac{ r^2}{\ell^2}
%\end{equation}
%The relation between the negative cosmological constant $\Lambda$ and the AdS radius $\ell$ is $\Lambda=-3/\ell^2$.
%
%
%+++++++Thus, t++++++++

%We consider the dynamics of a charged particle as it is absorbed by the  black hole. We concentrate mainly on the relations between the conserved quantities, such as the energy and the momentum. The Hamilton-Jacobi equation of the particle with the gauge potential $A_\mu$ is

The energy-momentum relation of a charged particle with the gauge potential $A_\mu$  near the event horizon as it's absorbed by the AdS black hole is unveiled by the dynamics of a charged scalar field under the Hamilton-Jacobi equation 
\begin{equation}\label{10}
g^{\mu \nu} \left( p_\mu - q A_\mu \right) \left( p_\nu - q A_\nu \right) = - m^2,
\end{equation}
where $m$ and $q$ are nothing than the rest mass and the electric charge of the particle respectively. Further, $p_\mu$ is the momentum of the particle, and $g^{\mu \nu}$ represents the metric tensor associated with a static spherical charged black hole in $4$-dimensional spacetime and which is given by
\begin{equation}\label{ad1}
g_{\mu \nu} dx^\mu dx^\nu = - f(r) dt^2 + \dfrac{dr^2}{f(r)}+ r^2 \left( d\theta^2 + \sin^2\theta d\varphi^2 \right) ,
 \end{equation}
 where the blackening function is 
\begin{equation}\label{ad2}
f(r) = 1 - \dfrac{2 M}{r} + \dfrac{Q^2}{r^{2}} + \dfrac{r^2}{\ell^2},
\end{equation}
with $\ell$ is the AdS radius. Hence, the energy of the particle absorbed by the black hole reads as
  \begin{equation}\label{19}
 E = \left| p^r_h\right| + \dfrac{q Q}{r_h},
 \end{equation}
 the positive sign in front of radial momentum $\left| p^r_h\right|$ term  assures the positive flow of time \cite{art6}. As the particle's energy \textbf{ $E$ is supposed always positive} \cite{Abbott:1981ff,Henneaux:1985tv,Aros:1999id,Chruciel2001TheMO, Maerten:2006eua,Chrusciel:2006zs,Wang,witten1981new, Nester:1981bjx,parker1982witten, Penrose:1993ud, Woolgar1994ThePO, GIBBONS1983173,Horowitz:1998ha, Galloway2003OnTG, Page2002PositiveMF}, the radial momentum $\left| p^r_h\right|$ should respect the following constraint 
   \begin{equation}\label{19_a}
  \left| p^r_h\right|>-\dfrac{q Q}{r_h} ,
 \end{equation}
 particularly in a neutralization process ($qQ<0$), the particle is attracted by the black hole because of the electric interactions.
 
 A black hole can change its state when it interacts with the external field and assumes that the final state of the black hole is still a black hole. The shift of $f(r)$ during such transformation is computed to be
 \begin{equation}\label{24}
df_h = d f (r_h) =\dfrac{\partial f_h}{\partial r_h} dr_h + \dfrac{\partial f_h}{\partial M} dM + \dfrac{\partial f_h}{\partial Q} dQ + \dfrac{\partial f_h}{\partial \ell} d\ell  = 0.
\end{equation}
At this level,  we are in a position to look for which thermodynamic potential is varied rigorously by the same quantity as that of the particle, namely its energy and charge. This is supported by the change of the black hole's state following the first law of thermodynamics which is given by Eq.\eqref{9}.
Concretely, we first consider the internal energy,  then the enthalpy, and check whether the second law persists or not. 

\begin{figure}[!ht]
			\centering \includegraphics[scale=0.6]{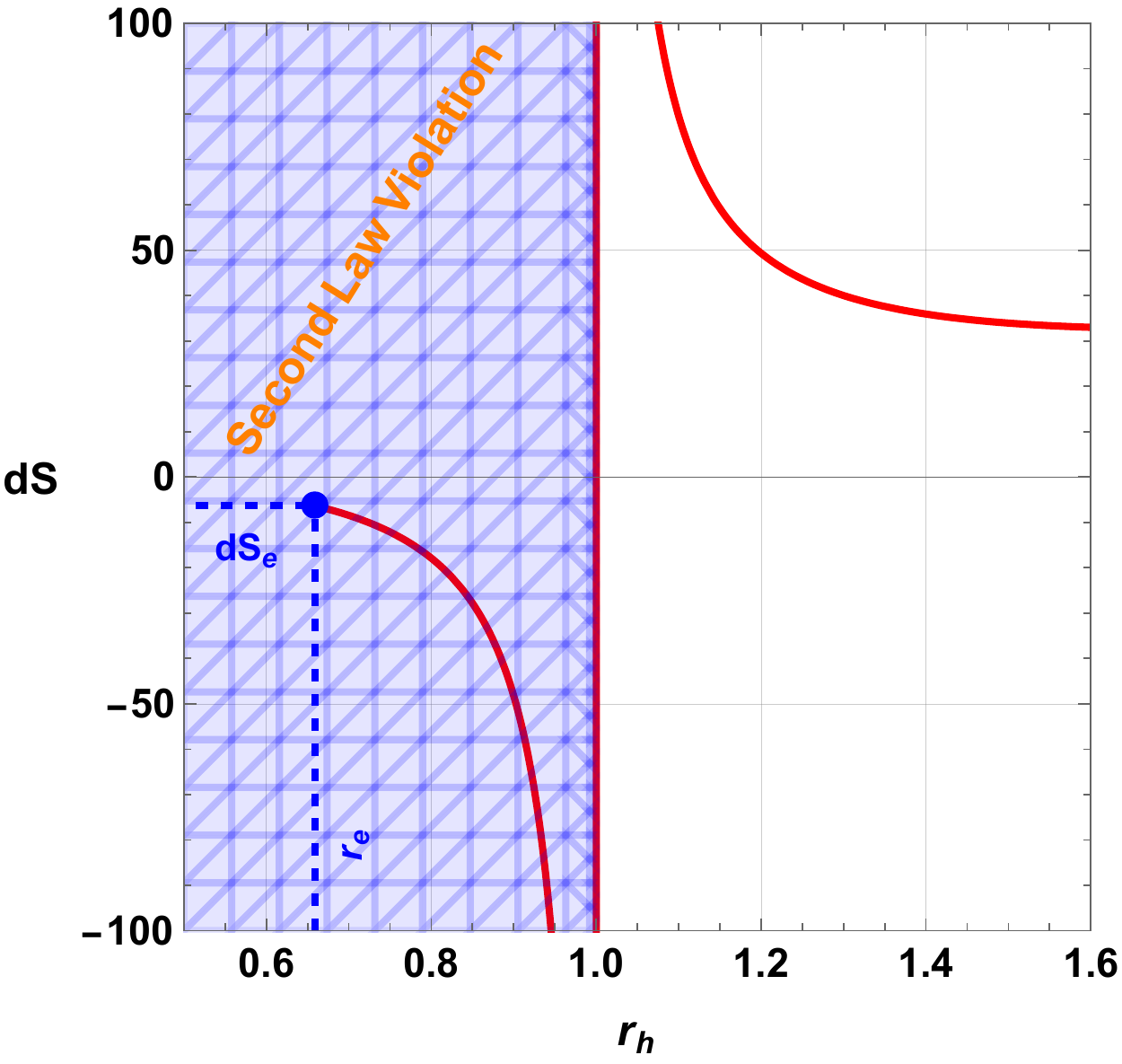}
	\caption{\footnotesize The entropy variation $dS$ as a function of event horizon radius $r_h$,   with $\left| p^r_h\right| = 1$, $Q=1$, $\ell=1$ and $dQ = 0.1$. }
	\label{f1}
\end{figure}

%\textcolor{red}{$\bigstar$}
\emph{Internal energy :}
we assume that the energy of the particle changes the internal energy of the black hole as it was supposed in \cite{Gwak:2019rcz}. The internal energy is given as $U\left( Q, S, V \right)  $, so it is a function of the charge, entropy, and the volume of the black hole. As the charged particle is absorbed by the black hole, the variation of internal energy should be a function of $dQ$, $dS$, and $dV$, which can be observed from the following equations
\begin{equation}\label{20}
E = dU = d\left( M-PV\right) , \quad \quad dQ = q.
\end{equation}
By the virtue of the first law of thermodynamics, one can write
\begin{equation}\label{21}
dU = T dS -  P dV + \Phi dQ= \left| p^r_h\right| + \dfrac{q Q}{r_h}.
\end{equation}
%As the charged particle is absorbed by the black hole, the variation of the entropy can be written as
In addition, the variation of the entropy can be expressed as
\begin{equation}\label{23}
dS = 2 \pi r_h d r_h ,
\end{equation}
where the change of the event horizon $dr_h$ should be rewritten as independent variables such as $(dQ, \left| p^r_h\right|)$ of the particle.  Knowing that the entropy and the volume depend only on the event horizon radius consequently their variations are proportional to $dr_h$, but the variation of internal energy depends differently on the variation of the volume and the entropy as according to Eq.\eqref{23}. Moreover, the first principle of thermodynamics Eq.\eqref{21} doesn't take into account the variation of the pressure $dP$, and the event horizon radius variation $d r_h$ is independent of $d\ell$. Thus increasing the internal energy does not imply the increase of the event horizon and hence the entropy. In order to verify this prediction, we calculate the variation of the event horizon resolving simultaneously Eq.\eqref{21} and Eq.\eqref{24}. Hence the variation of the event horizon  and thus the entropy variation can be found to be
\begin{equation}\label{26}
d r_h = \dfrac{2  \left| p^r_h\right| }{  \left( 1-\Phi^2\right) }\quad\Rightarrow\quad d S = \dfrac{4 \pi \left| p^r_h\right| r_h}{ \left( 1-\Phi^2\right) }.
\end{equation}
%\begin{equation}\label{27}
%d S = \dfrac{4 \pi \left| p^r_h\right| r_h}{ \left( 1-\Phi^2\right) }.
%\end{equation}
Obviously, the variation of the entropy could be negative when $\Phi^2 > 1 $. Thus we have a violation of the second law of thermodynamics when $r_h< r_0$ such that 
\begin{equation}\label{28}
r_0 = Q.
\end{equation}
We depict in Fig.\ref{f1} the behavior of the black hole entropy as a function of event horizon radius. We notice that the second law of thermodynamics is violated particularly near the extremal black hole where $dS_e=- \dfrac{4 \pi \left| p^r_h\right| \ell^2}{3 r_h  }< 0$ and for all black holes with $r_h<r_0$. Whereas, for $r_h>r_0$ the variation of the entropy is positive and the second law of thermodynamics is respected. %Moreover, the variation of the entropy diverge at $r_h = r_0$. Such behavior is unphysical because the black hole is regular where $r_h = r_0$. 
%"Therefore, we can easily conclude that the energy of the particle does not relate to the internal energy of the black hole."
The violation of the second law is most likely due to the incorrect assumption of the internal energy is changed by the infalling particle. Instead, as Ref.\cite{Hu:2019lcy} pointed out, it should be the enthalpy that is being changed. We should now turn to this assumption and examine it with greater detail beyond the analysis done in Ref.\cite{Hu:2019lcy}.

\emph{Enthalpy:} Now, we turn our attention to the second assumption in literature, where we assume that the energy of the particle changes the enthalpy $H$ of the black hole as it was supposed in \cite{Hu:2019lcy}. The enthalpy  is the Legendre transformation of the internal energy such as
\begin{equation}\label{30}
H = U+ PV.
\end{equation}
In this situation, the enthalpy $H$ corresponds to the mass of the black hole $M$ \cite{ar4,art5} and it's given as $H\left( Q, S, P \right)  $, so it is a function of the charge, entropy, and pressure of the black hole. As the charged particle is absorbed by the black hole, the variation of the enthalpy should be a function of $dQ$, $dS$, and $dP$, which can be interpreted by the following equation \begin{equation}\label{31}
E = dH = dM = T dS + V dP + \Phi dQ , \quad \quad dQ = q.
\end{equation}
Thus the variation of the enthalpy is given from Eq.\eqref{19}
\begin{equation}\label{32}
dM = \left| p^r_h\right| + \dfrac{q Q}{r_h}.
\end{equation}
According to Eq.\eqref{9}, the variation of the black hole mass can be now expressed as
\begin{equation}\label{33}
dM = \dfrac{\partial M}{\partial r_h} dr_h + \dfrac{\partial M}{\partial Q} dQ + \dfrac{\partial M}{\partial l} d\ell.
\end{equation}
Hence, the variation of the event horizon is given by
\begin{equation}\label{34}
dr_h = \dfrac{  2 \left[  \left| p^r_h\right| \ell^3 +  r_h^3  d\ell\right]  }{  \ell \left[ 3 r_h^2 +  \ell^2 \left( 1-\Phi^2\right) \right] }.
\end{equation}
Therefore, the variation of the black hole  entropy reads as
\begin{equation}\label{35}
d S = \dfrac{4 \pi r_h \left[  \left| p^r_h\right| \ell^3 +   r_h^3  d\ell\right]  }{  \ell \left[ 3 r_h^2 +  \ell^2 \left( 1-\Phi^2\right) \right] }.
\end{equation}
Herein, we are faced with a serious problem associated with a lack of information about the variation of pressure. 
 In other words, we are in possession of two unknown quantities, $dr_h$ and $d\ell$, and we have just one equation that comes from the first law of thermodynamics. Indeed, the equation Eq.\eqref{24} is equivalent to the first law equation and does not give us any further information. {\bf Hence, we grasp our need for another dynamical equation to close the equations system}. 

\emph{Grand potential:}
Statistical mechanics aims to provide a  microphysical underpinning for thermodynamics, justifying its foundational postulates and permitting derivations of its quantitative features from the microphysics of the system being studied.
Each thermodynamical system can be described entirely  by given the partition function which reads in the semiclassical limit 
\begin{equation}\label{36}
\mathcal{Z} = e^{- \mathcal{S}_E};
\end{equation}
where $\mathcal{S}_E$ is the euclidean action which is related to the free energy $F$ by %\footnote{In usual thermodynamical systems we have $F - N \mu = T \mathcal{S}_E$  where $F$ is Helmholtz energy, $N$ is the number of particles and $\mu$ is the chemical potential. In black hole thermodynamics we replace Helmholtz energy, $U-TS$, by Gibbs energy, $M -T S$, because the black hole mass plays the role of the enthalpy.}
\begin{equation}\label{37}
\Omega = F - Q \Phi = - T \log \left( \mathcal{Z}\right)  = T \mathcal{S}_E,
\end{equation}
such that $\Omega = F - Q \Phi $ is the grand potential \cite{art11, art12}. The least action principle in black hole thermodynamics stands for \cite{art13,art14,art15}, 
\begin{equation}\label{38}
d \Omega  = 0,
\end{equation}
which traduces the dynamical stability of the black hole and the grand potential should be minimized\footnote{$d \Omega  = E - d(TS)-d(Q\Phi)$, with $E$ is the energy of the absorbed particle.}. 

Now we have another equation besides the first law equation such that we can address our problem in a correct manner.  {\bf Resolving Eq.\eqref{32} and Eq.\eqref{38} simultaneously}, thus the variation of the event horizon and the AdS spacetime radius read as
\begin{equation}\label{39}
\begin{split}
dr_h &= E \\
d\ell &= -\dfrac{ \ell^3}{2 r_h} \left[ \dfrac{2 \left| p^r_h\right| + E \left( \Phi^2 -1\right)  }{ r_h^2} - \dfrac{3 E }{ \ell^2}\right] 
\end{split} .
\end{equation}   
Therefore, the variation of the black hole  entropy  can be expressed as
\begin{equation}\label{40}
 d S = 2 \pi E r_h, 
\end{equation}
and the variation of the pressure is given by 
\begin{equation}\label{42}
dP = - 2 P \dfrac{d\ell}{\ell}=\dfrac{2 \left| p^r_h\right| + E \left( \Phi^2 -1\right)  }{2 V} - \dfrac{3 P E }{r_h}.
\end{equation}

\begin{figure}[!th]
	\centering \includegraphics[scale=0.5]{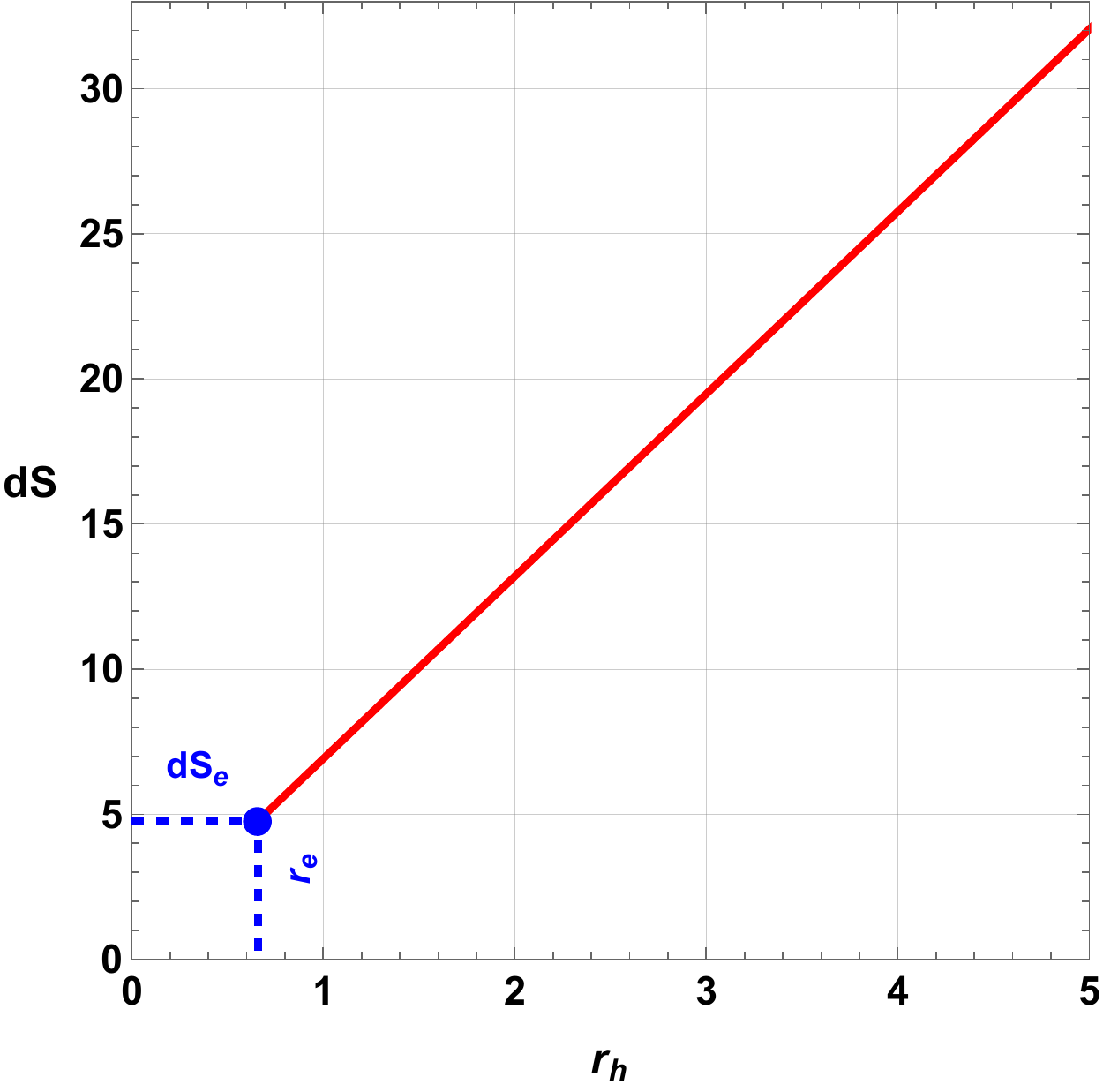}
		\caption{\footnotesize The entropy variation $dS$ as a function of event horizon radius $r_h$,  with $\left| p^r_h\right| = 1$, $Q=1$, $\ell=1$ and $dQ = 0.1$. }
	\label{f3}
\end{figure}

\begin{figure}[h!]
		\centering \includegraphics[scale=0.5]{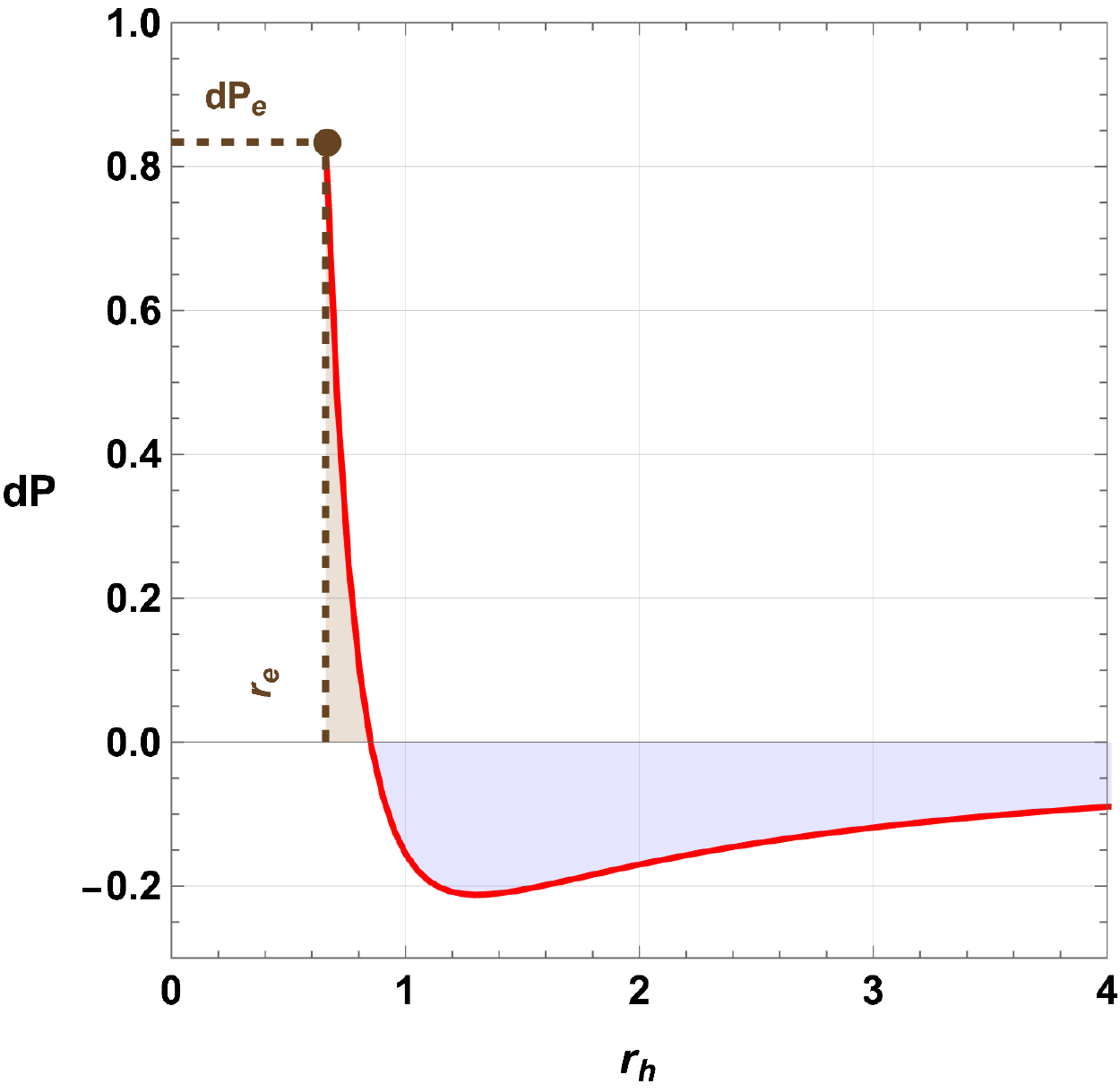}
			\caption{\footnotesize The pressure variation $dP$ as a function of event horizon radius $r_h$, with $\left| p^r_h\right| = 1$, $Q=1$, $\ell=1$ and $dQ = 0.1$. }
	\label{f4}
\end{figure}

We have, from Eq.\eqref{40}, the variation of the entropy is always positive and proportional to the event horizon radius.  We display in Fig.\ref{f3} the variation of the black hole entropy as a function of the event horizon radius and we observe that the variation of the entropy is positive and increases linearly when the black hole gets larger. 
We plot in Fig.\ref{f4} the variation of the $dP$ as a function of the event horizon radius $r_h$.  We observe that the variation of the pressure {\bf is no longer constant} as stipulated in \cite{Hu:2019lcy}.
Moreover, it can also change its sign, affecting the Eq.\eqref{35} if we taking it {\bf only}.
The main missing key in the evidence of no violation of the second law was taking into account the grand potential, which has led to a not complete understanding of the particle absorption by a black hole. Before concluding, it is important to highlight the recent work \cite{Dai:2021dog} in the same direction as our study, namely letting the cosmological constant vary dynamically to preserve the second law of thermodynamics and taking the environment of the black hole into account, but our study gives a complete and detailed picture of how the cosmological constant should vary using a thermodynamical description of the problem, whereas, in \cite{Dai:2021dog}, the authors have imposed a discrete variation of cosmological constant and have looked for the variation of the entropy in the function of the final black hole mass. 

%\section{Final Discussion}
To summarize, we have pointed out that the violation of the second law in the extended black hole thermodynamics was a result of incomplete assumptions that the energy of an infalling particle changes the internal energy of the black hole firstly, then, the enthalpy of the black hole that increases by the energy of the particle absorbed all in assuming a fixed CFT's number of $N$.  Whereas actually by introducing the grand potential we have managed to incorporate the missing pieces of information about the variation of the cosmological constant and thus the variation of $N$ in the second law.  %Therefore, no evidence of second law violation is no longer valid in the extended black hole thermodynamics. 
{\bf Therefore, the claim that the second law is violated in the extended black hole thermodynamics is in fact not valid.}
It was intriguing to show how this happens explicitly.

%
%
%Thus far, we have seen that some 
%
%Nevertheless, amongst the various possible Lorentzian analogs of the Thurston metrics, in this letter we study those found in
%
%Some also lead to applications in physics.  Recalling that
%
%statistical description of spacetime
%
%
%
%Maybe the unsatisfactory understanding of the
%
%
%We have shown that this concept could provide preliminary
%knowledge of no evidence of second law
%
%
%However, before pursuing this issue
%
%Fortunately, the well-known statistical mechanics

%\newpage
\bibliographystyle{apsrev4-1}%{rusnat}
\bibliography{biblio}
\end{document}